\documentclass[preprint,12pt]{elsarticle}
\usepackage{amssymb,amsmath,amscd}
\usepackage{amsfonts,bm,latexsym}
\usepackage{epsfig}
\usepackage{graphicx}
\usepackage{amsmath}
\usepackage{subfig}

\journal{Physics Letters A}

\begin{document}
\begin{frontmatter}

\title{Adiabatic sound velocity and compressibility of a trapped $d$-dimensional ideal anyon gas}

\author{Fang Qin\footnote{Email: qinfang.phy@gmail.com}
and Ji-sheng Chen\footnote{Email: chenjs@iopp.ccnu.edu.cn}}
\address{Physics Department and Institute of Nanoscience and
Nanotechnology, Central China Normal University, Wuhan 430079,
People's Republic of China}


\begin{abstract}

The adiabatic sound velocity and compressibility for harmonically trapped ideal anyons
in arbitrary dimensions are calculated within Haldane fractional exclusion statistics.
The corresponding low-temperature and high-temperature behaviors are studied in detail.
To compare with the experimental result of unitary fermions,
the sound velocity for anyons in the cigar-shaped trap is derived.
The sound velocity for anyons in the disk-shaped trap is also calculated.
With the parameter $g=0.287$, the sound velocity of unitary fermions in the
cigar-shaped trap modeled by anyons is in good agreement with the experimental result, while
that of unitary fermions in the disk-shaped trap is $v_{0}/v_{F}=0.406$ with Fermi velocity $v_{F}$.

\end{abstract}
\end{frontmatter}

\section{Introduction}\label{section2}

The statistical behaviors of a quantum many-body system can be successfully described by
Bose-Einstein or Fermi-Dirac statistics. Beside these two statistics, a new quantum statistics
called Haldane fractional exclusion statistics had also been proposed by Haldane \cite{Haldane1991}.

Following the distribution function of the Haldane anyon gas
given by Wu \cite{Wu1994}, the thermodynamics of ideal
anyons was a focus of theoretical attention in the past few years
\cite{Wu1994,Huang1996,Huang1998,Anghel2002,Qin1,Qin2,Qin3,Bhaduri1,Aoyama2001,Potter2007,Joyce1996,Nayak1994,Isakov1996,Khare1997,Iguchi1997,Sevincli}.
For example, the thermodynamics of the ideal anyons in two
dimensions has been investigated \cite{Huang1996,Huang1998,Anghel2002},
and the analytical expressions for the chemical potential, internal energy,
entropy, isochore and isobar heat capacities, and
Joule-Thomson coefficient of a three-dimensional
ideal anyon gas have been derived
in the previous works \cite{Qin1,Qin2,Qin3,Bhaduri1}.
Furthermore, the thermodynamics for a homogeneous ideal gas within
fractional exclusion statistics has been studied in some detail in
arbitrary dimensions by Refs.
\cite{Aoyama2001,Potter2007,Joyce1996,Nayak1994,Isakov1996,Khare1997,Iguchi1997}.
For a harmonically trapped $d$-dimensional ideal anyon system,
the internal energy and isochore heat capacity as a function of temperature
were given in Ref. \cite{Sevincli}.

In recent years, the strongly interacting physics in ultracold fermions
has attracted much attention both experimentally and theoretically
\cite{Giorgini2008,Bloch2008,Hu2007,Ho2004,Papenbrock2005,Qijin2,Luo2007,Joseph2007,Luo2009,Bulga2006}.
The ability to widely tune the effective interaction between two fermions by a broad Feshbach
resonance in the atoms of $^6$Li and $^{40}$K has permitted the
experimental observation of a smooth evolution of the Fermi
gas from the Bardeen-Cooper-Schrieffer (BCS) weakly attractive regime to
a molecular Bose-Einstein condensate (BEC) regime.
On the cusp of the BCS-BEC crossover, there is a strongly interacting
regime that is the so-called unitarity limit
regime \cite{Giorgini2008,Bloch2008}, which
corresponds to the unitary Fermi gas \cite{Hu2007}.
According to the theoretical universal hypothesis \cite{Ho2004}, the unitary
Fermi gas can exhibit a universal thermodynamic behavior.
The universality shows that the ground-state energy $\mu_{0}$ of a
homogeneous gas at zero temperature should be proportional to
the free Fermi energy $\epsilon_{F}$ with a proportionality constant $\xi$ ($\xi$=$1+\tilde{\beta}$)
which is the universal many-body constant.
The ratio of the ground-state energy to the Fermi energy for the unitary system
trapped in a harmonic trap
is $\xi^{1/2}$ with the density functional approach \cite{Papenbrock2005}.
Thus, $\mu_{0}=\xi^{1/2}\epsilon_{F}$, where $\epsilon_{F}$ is the trapped ideal Fermi energy.
Meanwhile, the thermodynamics of a trapped unitary Fermi gas has been experimentally measured
\cite{Qijin2,Luo2007,Joseph2007,Luo2009}.
Particularly, the ultracold sound velocity of a unitary Fermi gas
inside a strongly elongated cigar-shaped trap had been
measured by the Duke group \cite{Joseph2007,Luo2009}.
These recent experiments on the thermodynamics of ultracold Fermi gases provide an evidence
for the universality in the laboratory.
Furthermore, the important universal parameter $\xi$ (or $\tilde{\beta}$) can be extracted from
these experimental techniques and measured physical quantities.
In the measurement of sound velocity in a unitary Fermi gas,
Joseph et al. gave $\tilde{\beta}=-0.565\pm0.015$ \cite{Joseph2007}.

Physically, the thermodynamic properties of a unitary Fermi gas are between those of Fermi
and Bose gases \cite{Bulga2006}.
Meanwhile, the thermodynamic behaviors of the anyon gas obeying
the Haldane fractional exclusion statistics
are also between the fermionic and bosonic ones \cite{Wu1994}.
Due to the scale invariance, the thermodynamic quantities of the unitary Fermi gas and the Haldane anyon gas
are both related with the corresponding quantities of the ideal Fermi gas at zero temperature.
Therefore, it is assumed that the three-dimensional anyon gas obeying fractional exclusion
statistics can be used to model the statistical behavior
of a Fermi system at unitarity \cite{Qin1,Qin2,Qin3,Bhaduri1}.
In addition, the finite-temperature internal energy and entropy of the ideal anyons within
Haldane statistics are in good agreement with experimental data of the unitary fermions
for a given statistical parameter $g$ \cite{Qin2}.
Here, the value of $g$ is determined by the ground-state energy and
the value of the universal many-body constant $\xi$
extracted from the ultracold experiment.

In thermodynamics, the adiabatic sound velocity and compressibility for
the homogeneous quantum gases are two important quantities
\cite{Potter2007,Pathria1996,Potter20072,Chenjs,Liu,Zhang2011}.
The corresponding results
of the ideal Fermi and Bose gases were obtained by Ref. \cite{Pathria1996}.
The adiabatic sound velocity and compressibility of
a statistically interacting quantum gas in a generalized nonlinear
Schr{\"{o}}dinger model were calculated in Ref. \cite{Potter20072}. More
recently, for a strongly interacting Fermi gas, the adiabatic sound
velocity was discussed within the quasilinear approximation
framework \cite{Chenjs,Liu}, and it was also calculated with an
effective field theory near the Feshbach resonance regime
\cite{Zhang2011}.

In this Letter, the analytical expressions of the adiabatic sound velocity
and compressibility for the harmonically trapped ideal anyons in arbitrary
dimensions are derived within the Haldane fractional exclusion
statistics. The corresponding low-temperature and high-temperature
expansions of the thermodynamics are given.
We also investigate the zero-temperature sound velocities for
the three-dimensional anyon gases in the cigar-shaped
and disk-shaped harmonic traps to model the corresponding
physical quantities for a real unitary Fermi gas.

The outline is as follows. In Section
\ref{section2}, the grand canonical partition function of
the Haldane fractional exclusion statistics is given. The
adiabatic sound velocity and compressibility of a $d$-dimensional trapped
ideal anyon gas are derived analytically in Section \ref{section3}.
The corresponding numerical results are also evaluated in this section.
To compare with the recent experimental result of ultracold
fermions at unitarity, the zero-temperature sound velocity for a
three-dimensional anyon system
in a strongly elongated cigar-shaped harmonic trap
is derived in Section \ref{section4}.
In Section \ref{section5}, the sound velocity for a three-dimensional
anyon gas in a disk-shaped trap is also calculated analytically.
Summary and conclusions are presented in
Section \ref{section6}.

\section{Grand canonical partition function}\label{section2}

In the fractional exclusion statistics, the number of
microscopic quantum states of $N$ identical particles occupying
a group of $G$ states is \cite{Wu1994} \begin{eqnarray}
W=\prod_{i}\frac{[G_{i}+(N_{i}-1)(1-g)]!}{N_{i}![G_{i}-gN_{i}-(1-g)]!},
\end{eqnarray}
where the parameter $g$ connects the change
in the number of available states when one particle is added.

In order to find the grand canonical partition function, let us first focus
on the average occupation number $\langle n\rangle\equiv N_{i}/G_{i}$.
According to the Lagrange multiplier method
$\delta\ln{W}-\alpha\delta N-\beta\delta E=0$, the average occupation number
$\langle n\rangle$ is derived as \cite{Wu1994}
\begin{eqnarray}\label{f}
\langle n\rangle=\frac{1}{\omega+g},
\end{eqnarray} where $\omega$ and $g$ satisfy the relation
$\epsilon =k_{B}T\ln{\left[\omega^{g}(1+\omega)^{1-g}\right]}+\mu$
with the single-particle energy $\epsilon$, chemical potential
$\mu$, temperature $T$, and Boltzmann constant $k_{B}$.
$\alpha=-\mu/(k_{B}T)$ and $\beta=1/(k_{B}T)$ are the two Lagrange multipliers.
$N=\sum_{i}N_{i}$ is the total particle number and $E=\sum_{i}\epsilon_{i}N_{i}$
is the total energy. One has
$g=0$ for bosons and $g=1$ for fermions.
Analogous to the Fermi-Dirac statistics, the zero-temperature
average occupation number is $\langle n\rangle=0$ with $\epsilon>\mu$, and
$\langle n\rangle=1/g$ with $\epsilon<\mu$.

Setting the grand canonical partition function as
$\Xi=\prod_{\vec{p}}\Xi_{\vec{p}}$, the mean occupation number
$\langle n\rangle$ of momentum $\vec{p}$ turns out to be
\cite{Iguchi1997,Pathria1996}
\begin{eqnarray}\label{f2}
\langle n\rangle&&\equiv-\frac{1}{\beta}\left(\frac{\partial\ln{\Xi}}{\partial\epsilon}\right)_{z,T,all~other~\vec{p}}\nonumber\\
&&=-\frac{\omega(1+\omega)}{(\omega+g)}\frac{\partial\ln{\Xi_{\vec{p}}}}{\partial\omega},
\end{eqnarray} where the fugacity $z$ is defined as \cite{Qin1,Qin2,Qin3}
\begin{eqnarray}
z\equiv exp\left(\frac{\mu}{k_{B}T}\right)=\omega_{0}^{-g}(1+\omega_{0})^{g-1}.
\end{eqnarray}

Combining Eqs. (\ref{f}) and (\ref{f2}), one can conclude that
\begin{eqnarray}
\Xi_{\vec{p}}=1+\frac{1}{\omega},
\end{eqnarray} where the integral constant is set to be vanishing.
Correspondingly, the grand canonical partition function $\Xi$ can be identified to be
\begin{eqnarray}\label{X}
\Xi=\prod_{\vec{p}}\Xi_{\vec{p}}=\prod_{\vec{p}}\left(1+\frac{1}{\omega}\right).
\end{eqnarray}

\section{Adiabatic sound velocity and compressibility}\label{section3}

We now discuss the trapped system thermodynamics.
The geometric mean of the trap frequencies is defined as
$\varpi=\left(\prod_{i}\omega_{i}\right)^{1/d}$ ($i=1,2,\cdot\cdot\cdot,d$).
The density of states is
\begin{eqnarray}
D(\epsilon)=\frac{\epsilon^{d-1}}{(\hbar\varpi)^{d}\Gamma(d)},
\end{eqnarray} where $\hbar=h/(2\pi)$ is the reduced Planck constant,
and $\Gamma(d)$ is the gamma function.
The corresponding system volume is $V=\varpi^{-d}$ \cite{Sevincli,Romero2005}.
Notice that the volume here is no longer the usual thermodynamic
variable for a system in rigid walls without the
external potential. It is called the harmonic volume \cite{Romero2005}.
We will discuss the physical meaning of the harmonic volume
in the following of Eq. (\ref{S1}).

Utilizing Eq. (\ref{X}) and integrating by parts, the logarithm of the grand canonical
partition function $\Xi$ can be represented by turning the sum of quantum state into integral
\begin{eqnarray}\label{P1}
\ln\Xi&&=2\int_{0}^{\infty}{D(\epsilon)\ln{\left(1+\frac{1}{\omega}\right)}d\epsilon}\nonumber\\
&&=2\left(\frac{k_{B}T}{\hbar\varpi}\right)^{d}G_{d+1}(z,g),
\end{eqnarray} where
\begin{eqnarray}\label{G1}
G_{d}(z,g)=\frac{1}{\Gamma(d)}\int_{0}^{\infty}{\frac{x^{d-1}dx}{\omega+g}}
\end{eqnarray} is the Calogero-Sutherland integral function \cite{Potter2007}.
It satisfies the relation
\begin{eqnarray}\label{Gd}
z\frac{\partial G_{d}(z,g)}{\partial z}
=\begin{cases}
G_{d-1}(z,g), &\text{if $d\geqslant2$,} \\
\frac{1}{\omega_{0}+g}, &\text{if $d=1$.}
\end{cases}
\end{eqnarray}

Further, the grand thermodynamic potential is
$\Omega=-PV\equiv-k_{B}T\ln\Xi$, where $P$ is the press.
The expressions for the particle number $N$, and internal energy $E$ are
\begin{eqnarray}\label{N1}
N&&\equiv z\left(\frac{\partial\ln\Xi}{\partial z}\right)_{T,V}\nonumber\\
&&=2\left(\frac{k_{B}T}{\hbar\varpi}\right)^{d}G_{d}(z,g),
\end{eqnarray}
\begin{eqnarray}\label{E1}
E&&\equiv-\left(\frac{\partial\ln\Xi}{\partial\beta}\right)_{z,V}\nonumber\\
&&=2d\left(\frac{k_{B}T}{\hbar\varpi}\right)^{d}k_{B}TG_{d+1}(z,g).
\end{eqnarray}
From the thermodynamic formula $E-TS+PV\equiv N\mu$, one can
obtain the entropy per particle
\begin{eqnarray}\label{S1}
\frac{S}{N}=k_{B}\left[\frac{(d+1)G_{d+1}(z,g)}{G_{d}(z,g)}-\ln{z}\right].
\end{eqnarray}

Before making further analysis, let us discuss the
physical meaning of the harmonic volume $V=\varpi^{-d}$.
Since the temperature $T$ and chemical potential $\mu$ are intensive,
it is found that $\varpi^{-d}$ must be extensive
from the formulas of $\Omega$, $N$, $E$ and $S$.
At a given temperature, $\Omega$, $N$, $E$ and $S$ are all proportional to
$\varpi^{-d}$.
Thus, the harmonic volume $V=\varpi^{-d}$ does make physical sense as
a ``system volume" since small frequencies
of trap imply large actual volumes \cite{Romero2005}.
As shown in Eq. (\ref{N1}), $N\varpi^{d}\rightarrow$
constant in the thermodynamic limit, $N\rightarrow\infty$ and
$\varpi\rightarrow0$.
Therefore, the particle number density $n$ is set to be $n=N\varpi^{d}$
which will be used in the derivation of the adiabatic sound velocity below.

In terms of hydrodynamics, the expression for the sound velocity is
\begin{eqnarray}\label{v0}
v=\sqrt{\frac{1}{m}\left(\frac{\partial P}{\partial n}\right)_{S}}
\end{eqnarray} with the particle mass $m$.

According to the Jacobian techniques in thermodynamics, one has
\begin{eqnarray}\label{v1}
\left(\frac{\partial P}{\partial n}\right)_{S}
=\left(\frac{\partial(PV)}{\partial N}\right)_{T,V}
-\frac{\left(\frac{\partial(PV)}{\partial(k_{B}T)}\right
)_{N,V}\left(\frac{\partial(S/N)}{\partial
N}\right)_{T,V}}{\left(\frac{\partial
(S/N)}{\partial(k_{B}T)}\right)_{N,V}}.
\end{eqnarray}

With Eqs. (\ref{P1}), (\ref{Gd}), (\ref{N1}), (\ref{S1}),
(\ref{v0}), and (\ref{v1}), we finally obtain
the normalized sound velocity
\begin{eqnarray}\label{v2}
\frac{v}{v_{F}}=\sqrt{\frac{d+1}{2d}\frac{T}{T_{F}}\frac{G_{d+1}(z,g)}{G_{d}(z,g)}},
\end{eqnarray} where $v_{F}=\sqrt{2\epsilon_{F}/m}$ is the
zero-temperature Fermi sound velocity
of a harmonically trapped ideal Fermi gas.

As indicated in Eq. (\ref{v1}), we have transformed
the intensive thermodynamic quantizes $P$ and $n$ into the
extensive quantizes $\Omega=-PV$ and $N=n\varpi^{-d}$, which are not localized.
That is to say, $\Omega$ and $N$ are independent of location.
Essentially, the sound velocity obtained in Eq. (\ref{v2})
is an average velocity across the harmonic trap.
Thus, $v$ is independent of the position in the trap.

Similar to the derivation of the sound velocity, the
adiabatic compressibility turns out to be
\begin{eqnarray}\label{KS}
\kappa_{S}&&=-\frac{1}{V}\left(\frac{\partial V}{\partial P}\right)_{S}\nonumber\\
&&=\frac{d}{d+1}\frac{1}{nk_{B}T}\frac{G_{d}(z,g)}{G_{d+1}(z,g)}.
\end{eqnarray}

\subsection{Low-temperature behaviors}

In order to derive the low-temperature expressions of
the thermodynamic quantities, we consider that
the zero-temperature particle number is equal to
the finite-temperature one.

At zero temperature, it is worthy noting that the particle number
can be alternatively given by
\begin{eqnarray}\label{N0}
N=\frac{2}{g}\int_{0}^{\widetilde{\epsilon}_{F}}{D(\epsilon)d\epsilon}
=\frac{2\epsilon_{F}^{d}}{(\hbar\varpi)^{d}\Gamma(d+1)},
\end{eqnarray} where $\widetilde{\epsilon}_{F}$ obeys the relation
$\widetilde{\epsilon}_{F}=g^{{1}/{d}}\epsilon_{F}$ with the ideal
Fermi energy $\epsilon_{F}=[\Gamma(d+1)N/2]^{1/d}\hbar\varpi$
in a harmonic oscillator.

By substituting Eq. (\ref{N0}) into Eq. (\ref{N1}), one
gets
\begin{eqnarray}\label{N2}
\left(\frac{T}{T_{F}}\right)^{d}G_{d}(z,g)=\frac{1}{\Gamma(d+1)},
\end{eqnarray}
with the Fermi characteristic temperature
$T_{F}=[\Gamma(d+1)N/2]^{1/d}\hbar\varpi/k_{B}$
for a trapped ideal Fermi gas.

At very low temperature, the $G_{n}(z,g)$ can be expanded as
\cite{Potter2007,Joyce1996,Nayak1994,Isakov1996,Khare1997}
\begin{eqnarray}\label{GL}
G_{n}(z,g)=\frac{(\ln{z})^{n}}{g\Gamma(n+1)}
\left[1+\frac{\pi^2}{6}\frac{gn(n-1)}{(\ln{z})^{2}}+\cdot\cdot\cdot\right].
\end{eqnarray}

It should be pointed out that $G_{1}(z,g)$
can not be expanded by applying Eq. (\ref{GL}).
In fact, its expression with $\omega_{0}$ or temperature $T$ can be directly
given by Eqs. (\ref{G1}) and (\ref{N2})
\begin{eqnarray}\label{d1}
G_{1}(z,g)=\ln\left(\frac{1+\omega_{0}}{\omega_{0}}\right)=\frac{T_{F}}{T}.
\end{eqnarray}

Therefore, substituting Eq. (\ref{GL}) into (\ref{N2}) and (\ref{E1}), one
obtains the low-temperature analytical
expressions of the chemical potential and internal energy as follows:
\begin{eqnarray}\label{muL}
\frac{\mu}{\epsilon_{F}}=g^{1/d}
\left[1-\frac{(d-1)\pi^{2}}{6}g^{1-2/d}\left(\frac{T}{T_{F}}\right)^{2}+\cdot\cdot\cdot\right],
\end{eqnarray}
\begin{eqnarray}\label{EL}
\frac{E}{N\epsilon_{F}}=\frac{d}{d+1}g^{1/d}
\left[1+\frac{(d-1)\pi^{2}}{3}g^{1-2/d}\left(\frac{T}{T_{F}}\right)^{2}+\cdot\cdot\cdot\right].
\end{eqnarray}
For the special case $d=1$, the dominant asymptotic expansions
in Eqs. (\ref{muL}) and (\ref{EL})
vanish. Therefore, we obtain the very weak results
$\mu/\epsilon_{F}\sim g$ and $E/(N\epsilon_{F})\sim g/2$, as $T\rightarrow0$.

By applying Eqs. (\ref{GL}) and (\ref{muL}) to
Eqs. (\ref{S1}), (\ref{v2}), and (\ref{KS}), one has
\begin{eqnarray}\label{SL}
\frac{S}{Nk_{B}}=\frac{d\pi^{2}}{3}g^{1-1/d}
\left(\frac{T}{T_{F}}\right)+\cdot\cdot\cdot,
\end{eqnarray}
\begin{eqnarray}\label{vL}
\frac{v}{v_{F}}=\sqrt{\frac{g^{1/d}}{2d}}
\left[1+\frac{d\pi^{2}}{6}g^{1-2/d}\left(\frac{T}{T_{F}}\right)^{2}+\cdot\cdot\cdot\right],
\end{eqnarray}
\begin{eqnarray}\label{KSL}
\kappa_{S}=\frac{d}{ng^{1/d}\epsilon_{F}}
\left[1-\frac{(d+1)\pi^{2}}{6}g^{1-2/d}\left(\frac{T}{T_{F}}\right)^{2}+\cdot\cdot\cdot\right].
\end{eqnarray}

\subsection{High-temperature behaviors}

At very high temperature, the Calogero-Sutherland integral functions
have the power series expansions as follows \cite{Potter2007,Joyce1996}
\begin{eqnarray}\label{GH}
G_{n}(z,g)=z+\frac{z^{2}}{2^{n}}(1-2g)+\cdot\cdot\cdot.
\end{eqnarray}

Therefore, one gets
\begin{eqnarray}\label{vH}
\frac{v}{v_{F}}=\sqrt{\frac{d+1}{2d}\frac{T}{T_{F}}}
\left[1+\frac{z}{2^{d+2}}(2g-1)+\cdot\cdot\cdot\right],
\end{eqnarray}
\begin{eqnarray}\label{KSH}
\kappa_{S}=\frac{d}{d+1}\frac{1}{nk_{B}T}
\left[1+\frac{z}{2^{d+1}}(1-2g)+\cdot\cdot\cdot\right].
\end{eqnarray}
From Eqs. (\ref{vH}) and (\ref{KSH}), it is found that
$v/v_{F}\rightarrow\infty$ and $\kappa_{S}\rightarrow0$ when
$T/T_{F}\rightarrow\infty$.

\subsection{Numerical results and discussion}

Now, we will present the numerical results in this subsection.

\subsubsection{Adiabatic sound velocity}

The normalized adiabatic sound velocity for a trapped ideal anyon
system versus the rescaled temperature can be calculated from
Eqs. (\ref{v2}), (\ref{N2}), and (\ref{vL}).

\begin{figure}[htb]
  \centering
   \subfloat[]{\includegraphics[width = .45\textwidth]{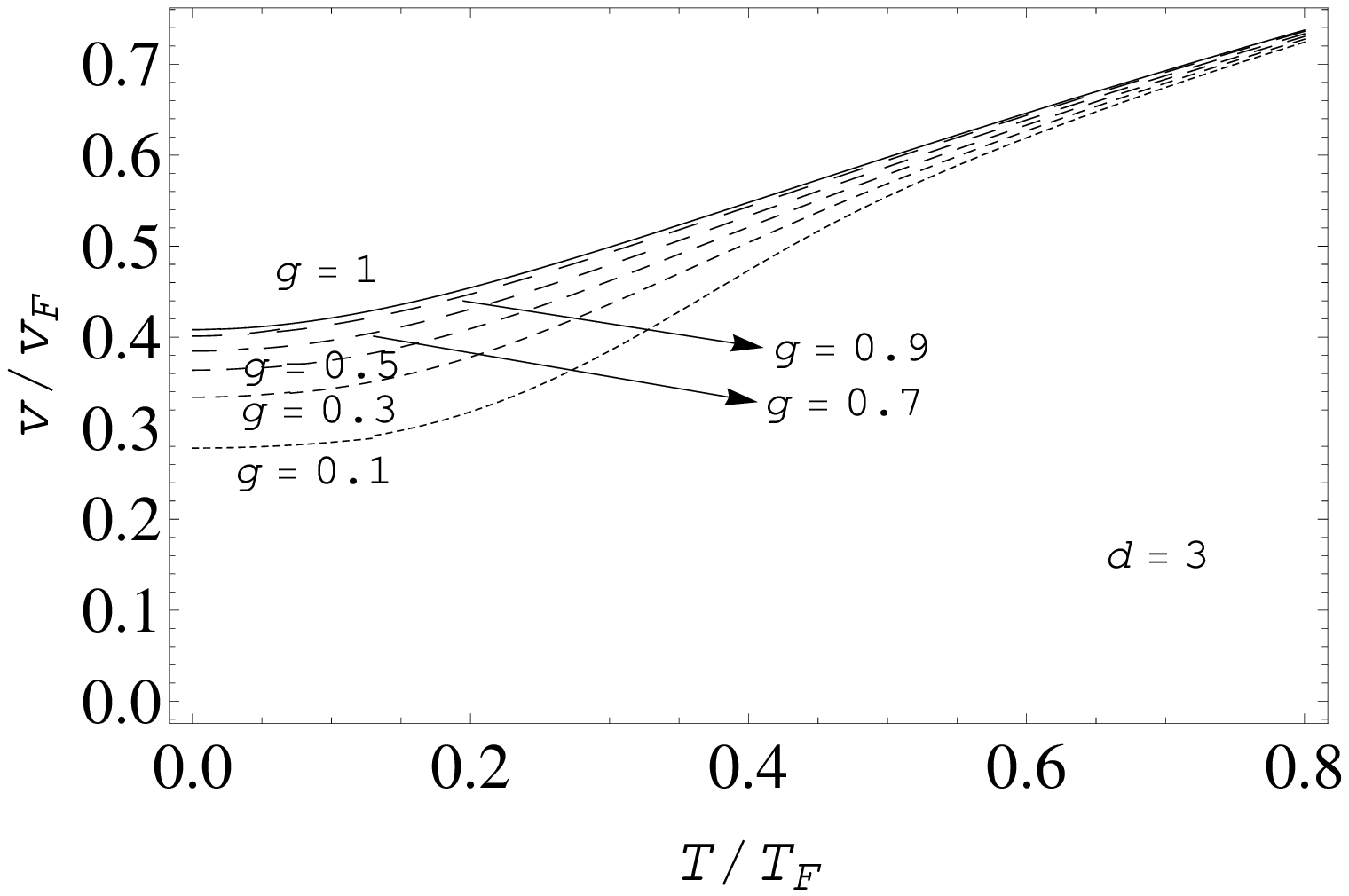}}
   \subfloat[]{\includegraphics[width = .45\textwidth]{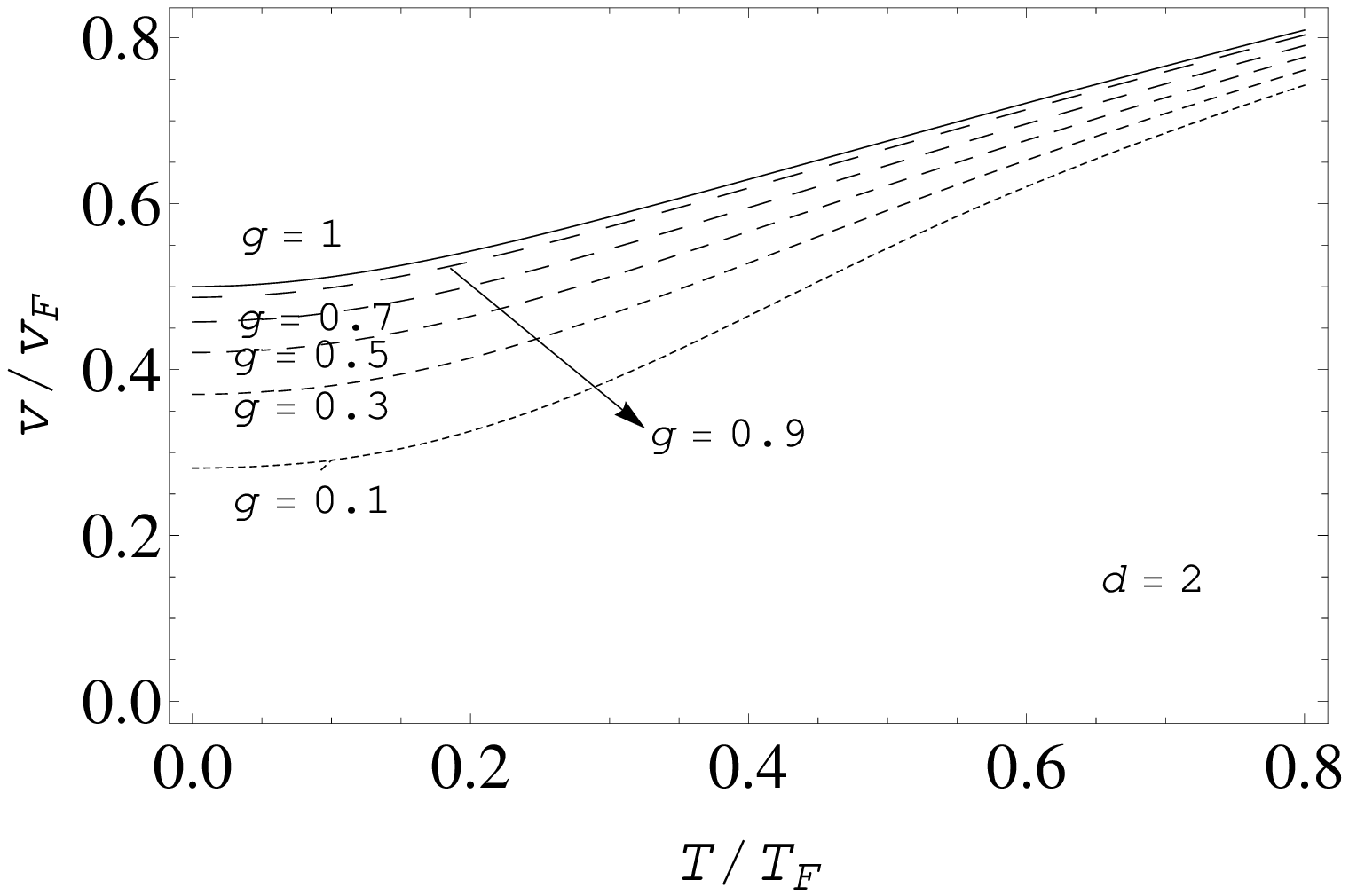}}\\
   \subfloat[]{\includegraphics[width = .45\textwidth]{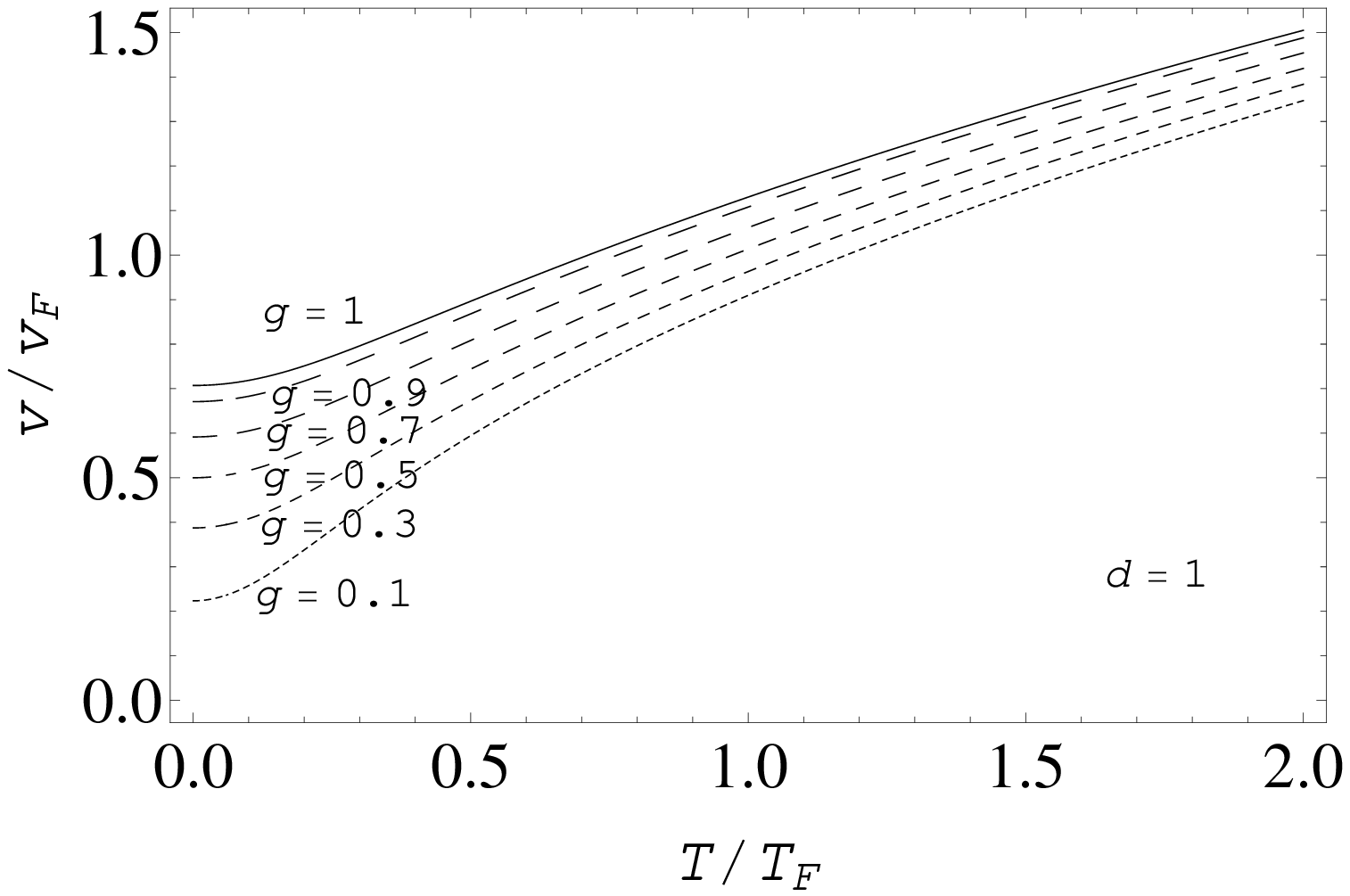}}
   \caption{\small
    The normalized adiabatic sound velocity is plotted as
    a function of the rescaled
    temperature ((a) $d=3$; (b) $d=2$; (c) $d=1$).
    The solid curve denotes
    that for the trapped ideal fermions, and the dashed curves denote
    the ones for trapped ideal anyons with statistical
    parameter $g=0.1,0.3,0.5,0.7,0.9$, respectively.}\label{fig1}
\end{figure}

As indicated by Fig. \ref{fig1},
the adiabatic sound velocity of a trapped ideal anyon gas increases
with the increase of the rescaled temperature. The curves for trapped anyons
get closer to that of the ideal fermions in the high-temperature
Boltzmann regime. The adiabatic sound velocity given by this model
is lower than the ideal fermionic one, and it goes to a smaller
value for a smaller value of $g$ at the same temperature.

\subsubsection{Adiabatic compressibility}

The adiabatic compressibility for a trapped ideal anyon
system versus the rescaled temperature can be calculated from
Eqs. (\ref{KS}), (\ref{N2}), and (\ref{KSL}).

\begin{figure}[htb]
  \centering
   \subfloat[]{\includegraphics[width = .45\textwidth]{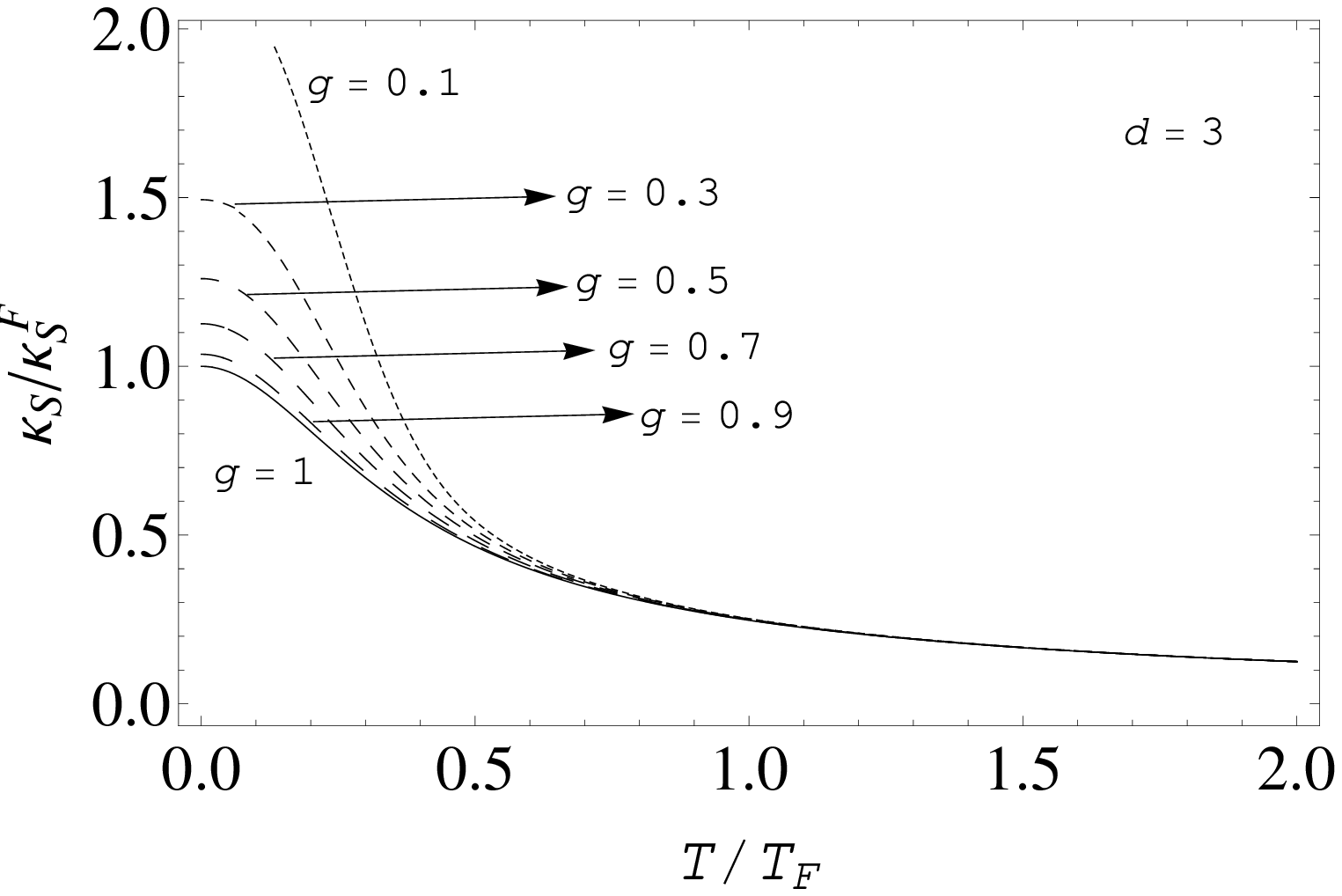}}
   \subfloat[]{\includegraphics[width = .45\textwidth]{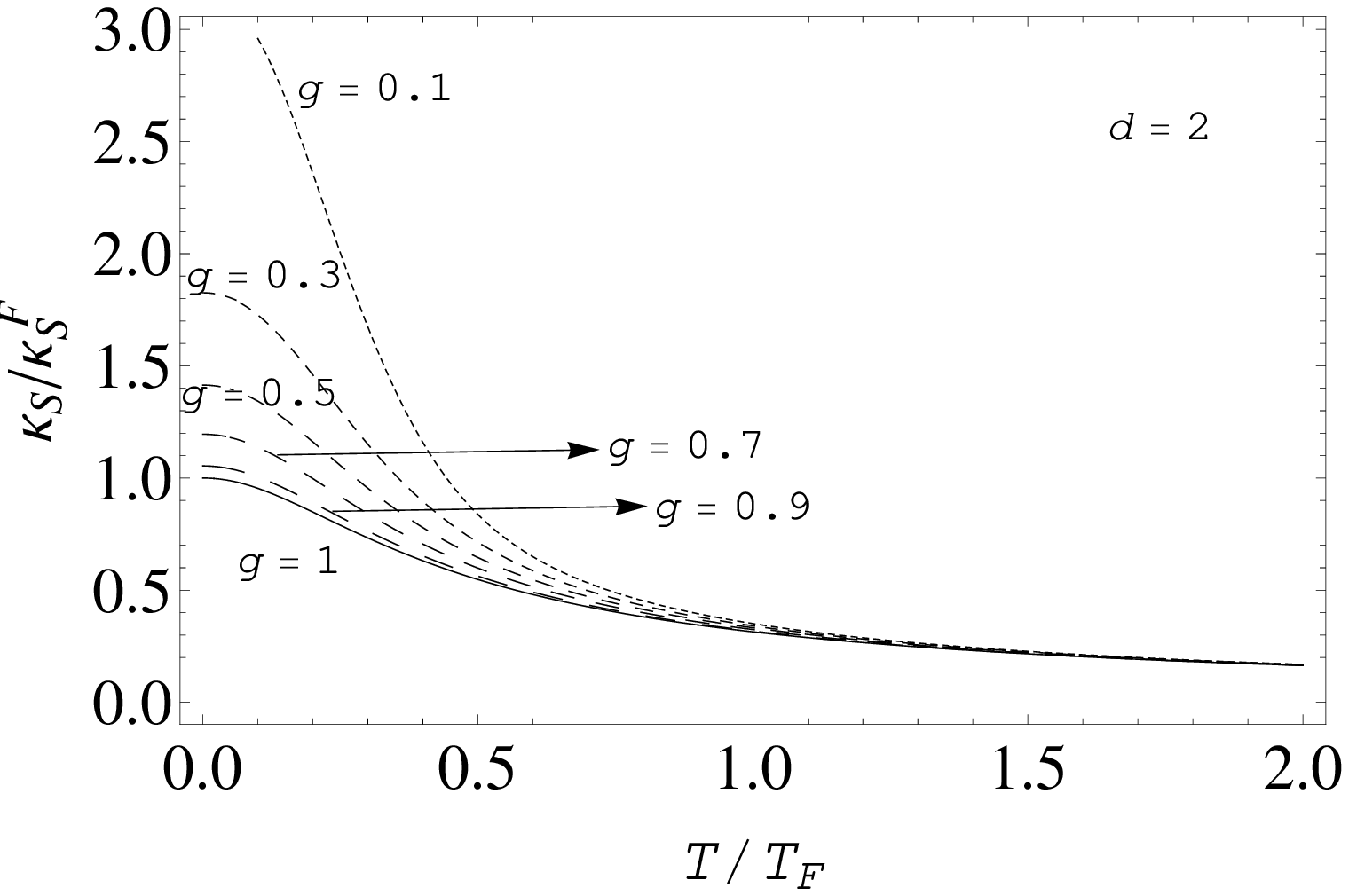}}\\
   \subfloat[]{\includegraphics[width = .45\textwidth]{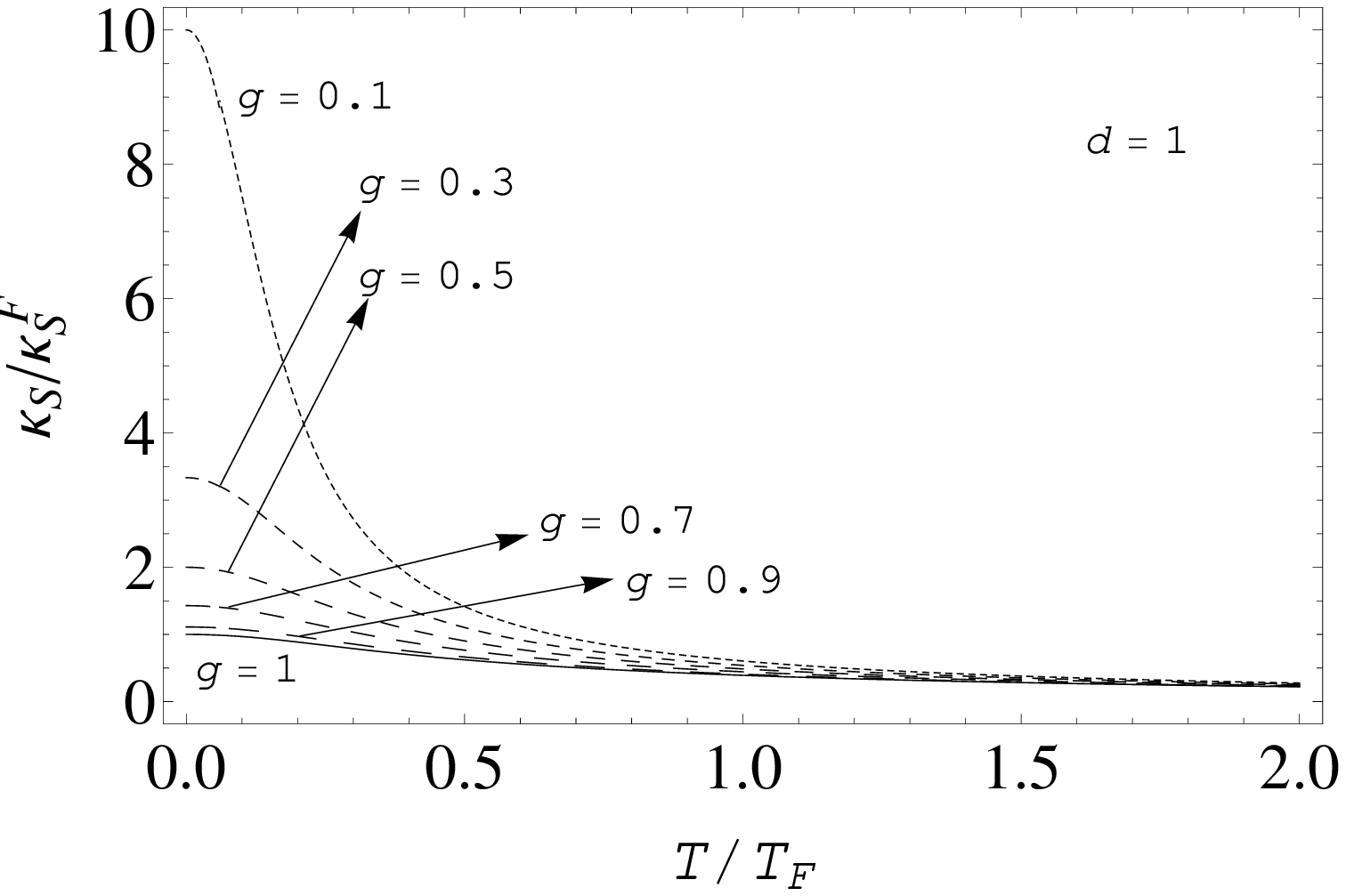}}
   \caption{\small
    The normalized adiabatic compressibility versus the rescaled
    temperature ((a) $d=3$; (b) $d=2$; (c) $d=1$).
    The line styles are similar to the ones in Fig. \ref{fig1}.
    It is indicated from Eq. (\ref{KSL}) that the
    zero-temperature compressibility is
    $\kappa_{S}^{F}=d/(n\epsilon_{F})$ for a trapped ideal Fermi gas.
    }\label{fig2}
\end{figure}

Fig. \ref{fig2} shows that
the adiabatic compressibility of a trapped ideal anyon gas decreases
with the increase of temperature. In the high-temperature limit,
the curves for trapped anyons get closer to that of the ideal fermions,
and they all tend to zero.
The adiabatic compressibility increases
with the decrease of $g$ for fixed $T$.

\section{Sound velocity for a three-dimensional anyon gas in a strongly elongated cigar-shaped trap}\label{section4}

Experimentally, the measurement of ultracold sound velocity for
a unitary Fermi gas in a strongly elongated cigar-shaped trap
had been given in Refs. \cite{Joseph2007,Luo2009}.
In order to compare with the experimental datum,
let us consider the case of a three-dimensional anyon system in a
strongly elongated cigar-shaped harmonic trap
$V(z,r_{\perp})=(m/2)(\omega_{z}^{2}z^{2}+\omega_{\perp}^{2}r_{\perp}^{2})$
with $\omega_{\perp}\gg\omega_{z}$. Here, $r_{\perp}$ is
the radial axis, $\omega_{\perp}$
is the radial trap frequency,
and $\omega_{z}$ is the axial trap frequency along $z$ axis.
For making an analytical approach available, we assume the trapping potential
is $z$ independent, i.e., $V(z,r_{\perp})=V(r_{\perp})=(m/2)\omega_{\perp}^{2}r_{\perp}^{2}$.

Within the local density approximation, the chemical potential of a
cigar-shaped harmonically trapped system in equilibrium can be expressed as
\begin{eqnarray}\label{rTF0}
\mu_{0}=\tilde{\mu}_{0}(r_{\perp})+\frac{m}{2}\omega_{\perp}^{2}r_{\perp}^{2},
\end{eqnarray} and the radial Thomas-Fermi radius $R_{\perp}$
is determined by
\begin{eqnarray}\label{rTF}
\mu_{0}=\frac{1}{2}m\omega_{\perp}^{2}R_{\perp}^{2},
\end{eqnarray} where $\mu_{0}$ is the zero-temperature chemical potential,
and $\tilde{\mu}_{0}(r_{\perp})$ is an effective chemical potential which
depends on $r_{\perp}$.

With Eqs. (\ref{rTF0}) and (\ref{rTF}), one can get
\begin{eqnarray}\label{nr}
n_{0}(r_{\perp})&&=n_{h}\left[\frac{\tilde{\mu}_{0}(r_{\perp})}{\mu_{0}}\right]^{3/2}\nonumber\\
&&=n_{h}\left[1-\left(\frac{r_{\perp}}{R_{\perp}}\right)^{2}\right]^{3/2},
\end{eqnarray}
where $n_{h}=(2m\mu_{0})^{3/2}/(3\pi^{2}\hbar^{3})$ that includes
two degrees of the spin degeneracy, and $R_{\perp}=[2\mu_{0}/(m\omega_{\perp}^{2})]^{1/2}$.
With the Thomas-Fermi approximation, the ground-state particle
number $N_{0}$ of the system is given by
\begin{eqnarray}\label{N0r}
N_{0}&&=\frac{1}{g}\int_{0}^{R_{\perp}}2\pi n_{0}(r_{\perp})r_{\perp}dr_{\perp}\int_{-Z/2}^{Z/2}{dz}\nonumber\\
&&=\frac{Z}{g}\frac{8(2m)^{1/2}}{15\pi\hbar^{3}\omega_{\perp}^{2}}\mu_{0}^{5/2}.
\end{eqnarray} If the cigar-shaped trap is equivalent to a cylinder, the
constant $Z$ in Eq. (\ref{N0r}) is the length of the cylinder.

Therefore, one can derive the corresponding normalized
sound velocity at zero temperature as
\begin{eqnarray}\label{v320}
\frac{v_{0}}{v_{F}}&&=\frac{1}{v_{F}}\sqrt{\frac{N_{0}}{m}\frac{\partial \mu_{0}}{\partial N_{0}}}\nonumber\\
&&=\sqrt{\frac{\mu_{0}}{5\epsilon_{F}}}.
\end{eqnarray}
For a harmonically trapped three-dimensional anyon gas at zero temperature, it
is indicated from Eq. (\ref{muL}) that
$\mu_{0}=g^{1/3}\epsilon_{F}$.
Correspondingly, with Eq. (\ref{v320}), one gets
\begin{eqnarray}\label{v321}
\frac{v_{0}}{v_{F}}=\frac{g^{1/6}}{\sqrt{5}}.
\end{eqnarray}
As discussed in the third paragraph of the introduction part, there is
$\mu_{0}=\xi^{1/2}\epsilon_{F}$ for a zero-temperature trapped unitary Fermi gas.
Comparing $\mu_{0}=g^{1/3}\epsilon_{F}$ with $\mu_{0}=\xi^{1/2}\epsilon_{F}$, one can find that
the relation between the statistical parameter $g$ and the universal constant $\xi$ is
$g=\xi^{3/2}$ \cite{Qin1,Qin2,Qin3}. Consequently, the final result is
\begin{eqnarray}\label{cigar-v}
\frac{v_{0}}{v_{F}}=\frac{\xi^{1/4}}{\sqrt{5}},
\end{eqnarray} which is consistent with Eq. (2) in Ref. \cite{Joseph2007}
based on the hydrodynamic approach that given by Capuzzi et al. \cite{Capuzzi2006}.

The recent experimental measurement on ultracold sound velocity for
a unitary Fermi gas in a strongly elongated cigar-shaped trap
provided $\xi=1+\tilde{\beta}=(1-0.565)\mp0.015$ \cite{Joseph2007}.
Therefore, one can determine the
statistical parameter $g=\xi^{3/2}=0.287$.

\section{Sound velocity for a three-dimensional anyon gas in a disk-shaped trap}\label{section5}

For the anyon gas in a disk-shaped harmonic trap with $\omega_{z}\gg\omega_{\perp}$,
we assume that $V(z,r_{\perp})=V(z)=(m/2)\omega_{z}^{2}z^{2}$.
The chemical potential is
$\mu_{0}=\tilde{\mu}_{0}(z)+(m/2)\omega_{z}^{2}z^{2}$, and
the axial Thomas-Fermi radius $R_{z}$ is defined as
$R_{z}=(1/\omega_{z})\sqrt{2\mu_{0}/m}$.

The corresponding ground-state particle number
takes the following form
\begin{eqnarray}\label{N0z}
N_{0}&&=\frac{1}{g}\int_{-R_{z}}^{R_{z}}n_{0}(z)dz\int_{0}^{R}2\pi r_{\perp}{dr_{\perp}}\nonumber\\
&&=\frac{\pi R^{2}}{g}\frac{m}{2\pi\hbar^{3}\omega_{z}}\mu_{0}^{2},
\end{eqnarray} where
\begin{eqnarray}\label{nz}
n_{0}(z)=n_{h}\left[1-\left(\frac{z}{R_{z}}\right)^{2}\right]^{3/2}.
\end{eqnarray} If the disk-shaped trap is regarded as a cylinder, the
constant $\pi R^{2}$ in Eq. (\ref{N0z}) is the cross-sectional area of the cylinder.

Similarly, the reduced zero-temperature sound velocity is given by
\begin{eqnarray}\label{v320z}
\frac{v_{0}}{v_{F}}=\frac{1}{2}\sqrt{\frac{\mu_{0}}{\epsilon_{F}}}.
\end{eqnarray}
Substituting $\mu_{0}=g^{1/3}\epsilon_{F}$ and $g=\xi^{3/2}$
into Eq. (\ref{v320z}), one obtains
\begin{eqnarray}\label{disk-v}
\frac{v_{0}}{v_{F}}=\frac{\xi^{1/4}}{2}.
\end{eqnarray}
With $g=0.287$,
the normalized sound velocity of a unitary Fermi gas
in a disk-shaped trap is $v_{0}/v_{F}=0.406$.

\section{Summary and conclusions}\label{section6}

The analytical expressions of the adiabatic sound velocity
and compressibility for a harmonically trapped
ideal anyon gas in arbitrary dimensions are derived
within the Haldane fractional exclusion statistics.
The corresponding low-temperature and
high-temperature expansions are also studied.

With careful numerical study, we find that
the adiabatic sound velocity of a trapped ideal anyon gas is a monotone
increasing function of temperature for a given $g$, and
the curves for the trapped anyon
gases get closer to that of the ideal fermions in the
high-temperature Boltzmann regime. The adiabatic sound velocity
given by this model is lower than the ideal fermionic one, and it
goes to a larger value for a larger value of $g$ at the same
temperature. However,
the adiabatic compressibility is a monotone
decreasing function of temperature for a fixed $g$, and it
goes to a larger value for a smaller value of $g$ at the same
temperature. The adiabatic compressibility
curves all tend to zero in the high temperature limit.

The ultracold sound velocity for a three-dimensional
unitary Fermi gas in a strongly elongated
cigar-shaped trap modeled by anyon statistics
is consistent with the experimental measurement for a given $g$.
The sound velocity for a three-dimensional anyon gas in a
disk-shaped trap is also obtained.
With $g=0.287$, the normalized sound velocity
of a unitary Fermi gas in the disk-shaped trap is $v_{0}/v_{F}=0.406$.

\section*{Acknowledgements}
This work was supported by the National Natural
Science Foundation of China under Grant Nos. 10875050 and
11178001.


\begin{thebibliography}{99}

\bibitem{Haldane1991} F. D. M. Haldane, Phys. Rev. Lett. \textbf{67} (1991) 937.

\bibitem{Wu1994} Y.-S. Wu, Phys. Rev. Lett. \textbf{73} (1994) 922;
Phys. Rev. Lett. \textbf{74} (1995) 3906.

\bibitem{Huang1996} W.-H. Huang, Phys. Rev. \textbf{B 53} (1996) 15842.

\bibitem{Huang1998} W.-H. Huang, Phys. Rev. Lett. \textbf{81} (1998) 2392.

\bibitem{Anghel2002} D. V. Anghel, J. Phys. \textbf{A 35} (2002) 7255.

\bibitem{Qin1} F. Qin and J.-S. Chen, Phys. Rev. \textbf{A 79} (2009) 043625.

\bibitem{Qin2} F. Qin and J.-S. Chen, J. Phys. \textbf{B 43} (2010) 055302.

\bibitem{Qin3} F. Qin and J.-S. Chen, Phys. Rev. \textbf{E 83} (2011) 021111.

\bibitem{Bhaduri1} R. K. Bhaduri, M. V. N. Murthy, and M. K. Srivastava, J. Phys. \textbf{B 40} (2007) 1775.

\bibitem{Aoyama2001} T. Aoyama, Eur. Phys. J. \textbf{B 20} (2001) 123.

\bibitem{Potter2007} G. G. Potter, G. M\"{u}ller, and M. Karbach, Phys. Rev. \textbf{E 75} (2007) 061120.

\bibitem{Joyce1996} G. S. Joyce, S. Sarkar, J. Spalek, and K. Byczuk, Phys. Rev. \textbf{B 53} (1996) 990.

\bibitem{Nayak1994} C. Nayak and F. Wilczek, Phys. Rev. Lett. \textbf{73} (1994) 2740.

\bibitem{Isakov1996} S. B. Isakov, D. P. Arovas, J. Myrheim, and A. P. Polychronakos, Phys.
Lett. \textbf{A 212} (1996) 299.

\bibitem{Khare1997}
A. Khare, \textit{Fractional Statistics and Quantum Theory}, (World
Scientific, Singapore, 1997).

\bibitem{Iguchi1997} K. Iguchi, Phys. Rev. Lett. \textbf{78} (1997) 3233.

\bibitem{Sevincli} S. Sevin\c{c}li and B. Tanatar, Phys. Lett. \textbf{A 371} (2007) 389.

\bibitem{Giorgini2008} S. Giorgini, L. P. Pitaevskii, and S. Stringari, Rev. Mod. Phys. \textbf{80} (2008) 1215.

\bibitem{Bloch2008} I. Bloch, J. Dalibard, and W. Zwerger, Rev. Mod. Phys. \textbf{80} (2008) 885.

\bibitem{Hu2007} H. Hu, P. D. Drummond and X.-J. Liu, Nature Physics \textbf{3} (2007) 469.

\bibitem{Ho2004} T.-L. Ho, Phys. Rev. Lett. \textbf{92} (2004) 090402.

\bibitem{Papenbrock2005} T. Papenbrock, Phys. Rev. \textbf{A 72} (2005) 041603(R).

\bibitem{Qijin2} J. Kinast, A. Turlapov, J. E. Thomas, Q. Chen, J. Stajic and K.
Levin, Science \textbf{307} (2005) 1296.

\bibitem{Luo2007} L. Luo, B. Clancy, J. Joseph, J. Kinast and J. E. Thomas, Phys.
Rev. Lett. \textbf{98} (2007) 080402.

\bibitem{Joseph2007} J. Joseph, B. Clancy, L. Luo, J. Kinast, A. Turlapov, and J. E. Thomas, Phys. Rev. Lett. \textbf{98} (2007) 170401.

\bibitem{Luo2009} L. Luo and J. E. Thomas, J. Low. Temp. Phys. \textbf{154} (2009) 1.

\bibitem{Bulga2006} A. Bulgac, J. E. Drut, and P. Magierski, Phys. Rev. Lett. \textbf{96} (2006) 090404.

\bibitem{Pathria1996} R. K. Pathria, \textit{Statistical Mechanics},
2nd edition, (Butterworth-Heinemann, Oxford, 1996).

\bibitem{Potter20072} G. G. Potter, G. M\"{u}ller, and M. Karbach, Phys. Rev. \textbf{E 76} (2007) 061112.

\bibitem{Chenjs} J.-S. Chen, J. Stat. Mech.: Theory Exp. L08002 (2009).

\bibitem{Liu} K. Liu and J.-S. Chen, Chin. Phys. \textbf{B 20} (2011) 020501.

\bibitem{Zhang2011} Z. Zhang and W. V. Liu, Phys. Rev. \textbf{A 83} (2011) 023617.

\bibitem{Romero2005} V. Romero-Roch\'{i}n, Phys. Rev. Lett. \textbf{94} (2005) 130601.

\bibitem{Capuzzi2006} P. Capuzzi, P. Vignolo, F. Federici, and M. P. Tosi, Phys. Rev. \textbf{A
73} (2006) 021603(R).

\end{thebibliography}
\end{document}